\documentclass[showpacs, prl, aps, superscriptaddress, twocolumn]{revtex4}
\usepackage{mathrsfs}
\usepackage{array}
\usepackage{color}
\usepackage{amsmath}
\usepackage{graphicx}
\usepackage{amstext}
\usepackage{amsfonts}
\usepackage{bm}
\usepackage{epstopdf}

\begin{document}
\title{Decoherence dynamics of Majorana qubits under braiding operations}
\author{Hon-Lam Lai}
\affiliation{Department of Physics and Center for Quantum
information Science, National Cheng Kung University, Tainan 70101,
Taiwan}
\author{Wei-Min Zhang}
\email{wzhang@mail.ncku.edu.tw}
\affiliation{Department of Physics
and Center for Quantum information Science, National Cheng Kung
University, Tainan 70101, Taiwan}
\begin{abstract}
We study the decoherence dynamics of Majorana qubit braiding operations in a topological superconducting chain (TSC) system, in which the braiding is performed
by controlling the electron-chemical potentials of the TSCs and the couplings between them. By solving rigorously the Majorana
qubit dynamics, we show how the Majorana qubit coherence is generated through bogoliubon correlations formed by exchanging Majorana zero modes (MZMs)
between two TSCs in braiding operations. Using the exact master equation, we demonstrate  how MZMs and also the bogoliubon correlations
dissipate due to charge fluctuations of the controlling gates at both the zero and finite temperatures.
As a result, Majorana qubit coherence and the fermion parity conservation cannot be immune from local perturbations during braiding operations.
\end{abstract}

\pacs{72.10.Bg, 73.63.-b, 85.30.Mn} \maketitle

%\section{Introduction}
%\noindent {\it Introduction}.
Topological quantum computation (TQC) has recently attracted lots of attention as a promising candidate for the realization of quantum computation.
Topological qubits consist of non-local degenerate ground states of Bogoliubov quasiparticles (bogoliubons) in topological superconductors, known as
Majorana zero modes (MZMs) [\onlinecite{Kitaev2003}-\onlinecite{RevPhys2008}], and qubit operations are performed by braidings between
MZMs [\onlinecite{Nature2011}-\onlinecite{meas_braid3}]. It has been commonly believed that
topological qubits have two main advantages [\onlinecite{Kitaev2001}]:
(1) the qubit is protected by the fermion parity conservation in superconducting systems as long as perturbations are small compared
to the energy gap between the MZMs and the non-zero energy bogoliubon states;
(2) the two MZMs of each zero-energy bogoliubon are non-locally separated from each other so that
topological qubits are considered to be immune to local perturbations during qubit manipulations.

However, the above picture of MZMs being robust against decoherence is only an ideal thought. In reality, performing braidings of MZMs
inevitably leads to couplings between the MZMs and the controlling gates. In other words, topological qubits must be influenced from
various environment noises, and the above two advantages of topological qubits should be reexamined in the open quantum
system perspective. It is easy to understand that coupling MZMs to an ungapped fermionic bath could break
the fermion parity conservation because electrons can tunnel into the bath without energy cost [\onlinecite{Budich2012}].
Even when the MZMs is coupled to a gapped fermionic bath through a bosonic field, electrons can still overcome the energy gap
and tunnel to the bath due to the bosonic field fluctuation [\onlinecite{Goldstein2011}, \onlinecite{Daniel2012}]. In our recent work,
we applied our exact master equation to a TSC system subjected to charge fluctuations of a controlling gate and studied the exact
non-Marokovian decoherence dynamics of MZMs [\onlinecite{PRB2018}]. We found that at zero temperature, there is a localized bound
state located at zero energy due to the particle-hole symmetry that can partially protect the MZM from decoherence. At any low
but finite temperature, MZMs can be locally destroyed completely.

We shall study in this Letter the non-local feature of the MZMs through the braiding operations, which is crucial for the realization of topological
quantum computers. Qubit information cannot be encoded in the two degenerate states
of one single zero-energy bogoliubon due to
the parity conservation. Therefore, at least four MZMs (two bogoliubons) must be involved in encoding quantum information.
%Each qubit consists of two non-local MZMs with the same parity from the two bogoliubons. ???
More specifically, two bogoliubons correlate to each other to encode non-locally qubit information
in the same parity sector through the braiding operation of MZMs.
We find that local perturbations to a single MZM can destroy the correlation between different
bogoliubons and thus destroy the quantum information encoded in the topological qubit. Therefore, even though local perturbations
to a single MZM cannot affect the wavefunction of another spatially separated MZM of the same bogoliubon, this does not
imply the protection of  topological qubit information from local perturbations. In fact, as we shall show in this Letter, Majorana qubit coherence is
generated through the formation of bogoliubon correlation between different TSCs, and local perturbations that destroy bogoliubon
correlations will naturally destroy the Majorana qubit coherence in topological quantum computations.

Because of the close relation between bogoliubon correlations and qubit coherence, we study explicitly Majorana qubit decoherence
through the exploration of the decoherence dynamics of bogoliubon correlations in braiding operations.
%There are several works that studied the decoherence dynamics of Majorana qubits [], most of them are under certain
%approximations and none of them focus on the bogoliubon correlations in a concrete braiding process. In the present work,
More specifically, we study the Majorana qubit decoherence dynamics under direct braiding of MZMs in a quasi-1D
network with the exact master equation in this Letter. Braidings of MZMs in quasi-1D networks can be realized by tuning the chemical potentials of TSCs
and the coupling potentials between them. The chemical potentials of TSCs can be tuned by ``keyboard'' gates that individually
control the electron-chemical potentials at different parts of the network [\onlinecite{Nature2011}]. The coupling potentials
between different TSCs can be tuned by gates that control the tunneling barriers between MZMs [\onlinecite{coup_braid}-\onlinecite{coup_braid2}].
We treat the T-junction-TSC system as an open quantum system, in which the MZMs can be moved by braiding through
controlling gates. We then solve the Majorana correlation functions and the reduced density matrix of the Majorana qubit
in the real-time domain in which all the non-Markovian effects are fully taken into account, and from which we explore explicitly
how braiding operations of MZMs generates Majorana qubit coherence and how non-adiabatic braiding operation and the
associated charge-fluctuation induced decoherence take place.

%\section{Majorana braiding and non-adiabatic decoherence dynamics}
\hspace{0.05cm}

\begin{figure}
\centerline{\scalebox{0.26}{\includegraphics{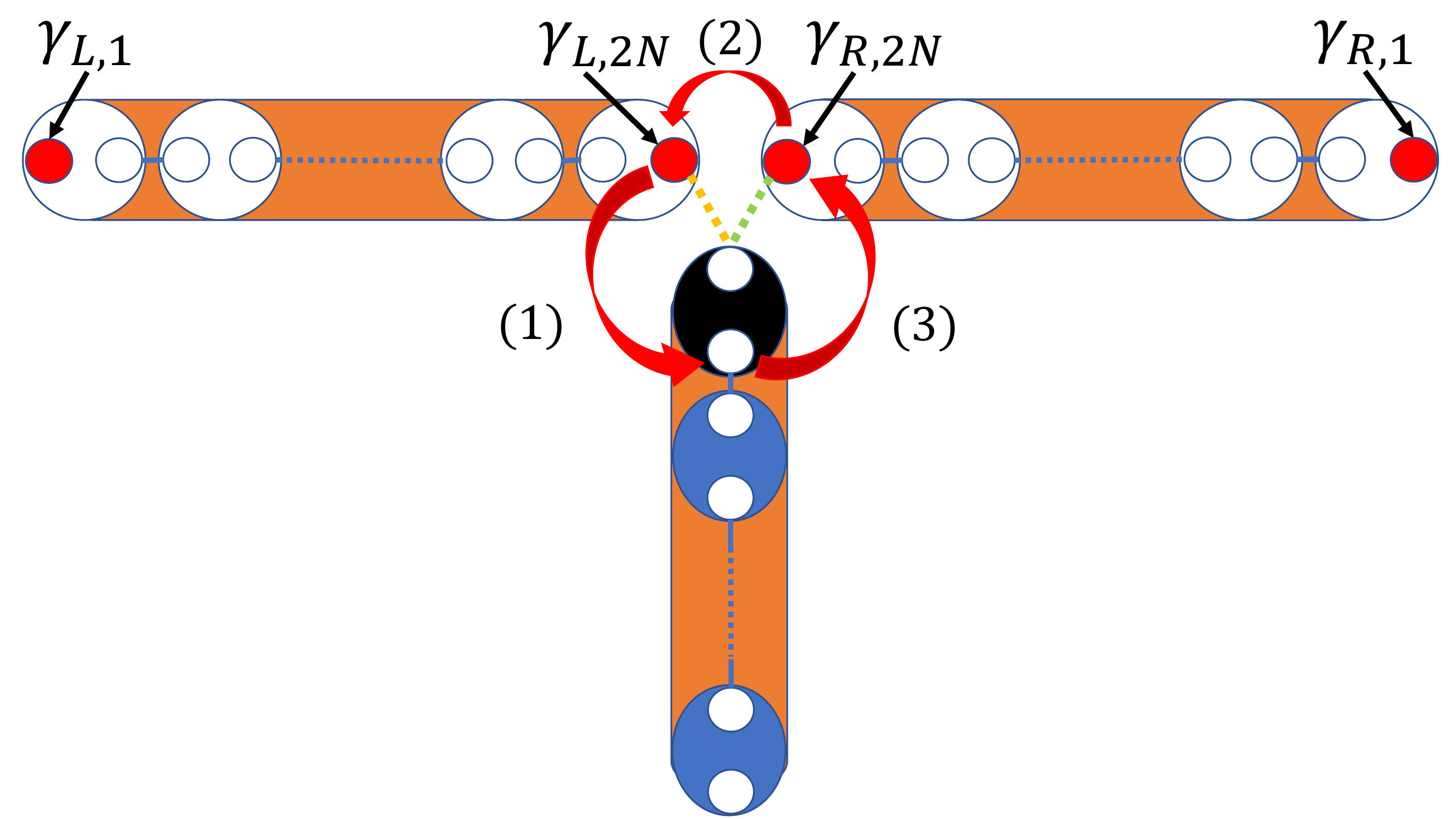}}}
\centerline{\scalebox{0.26}{\includegraphics{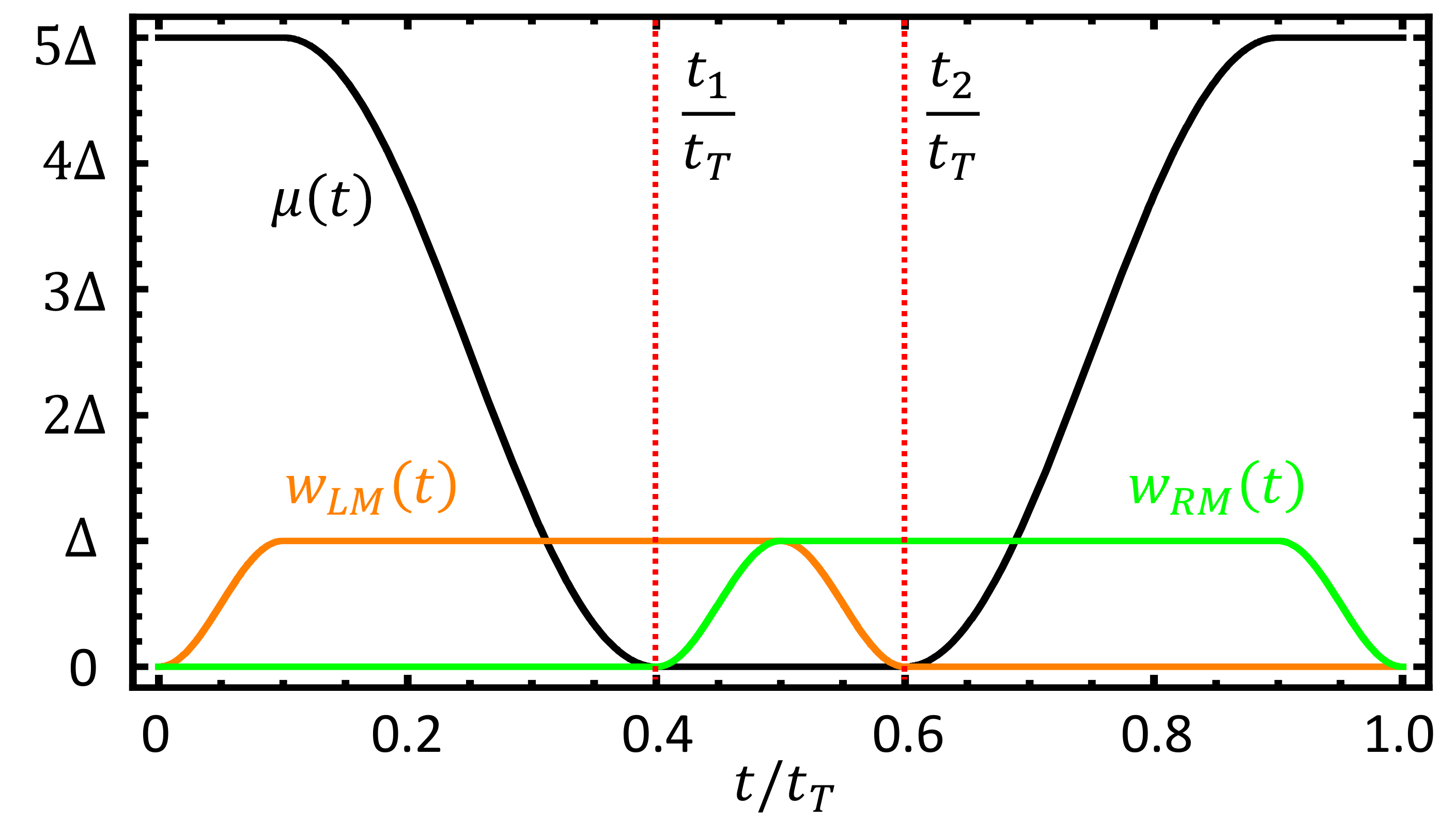}}}
\caption{(Colour online) (a) (Top) Three TSCs form a T-junction with four MZMs at sites $\gamma_{L,1}$, $\gamma_{L,2N}$, $\gamma_{R,1}$ and $\gamma_{R,2N}$ respectively.
Braiding operation is performed in three steps (1), (2) and (3) by tuning the chemical potential
$\mu(t)$ of the middle TSC and the couplings $w_{LM}(t)$ and $w_{RM}(t)$ of the left TSC and the right TSC with
the middle TSC, respectively. (b) (Bottom) Changes of parameters in a braiding
process: $\mu_{M,1}(t)$ (black), $w_{LM}(t)$ (orange) and $w_{RM}(t)$ (green).}\label{fig1}
\end{figure}

\noindent {\it Majorana braiding and non-adiabatic decoherence dynamics}.
We begin with a T-junction consisting of three TSCs (see Fig.~\ref{fig1}), each TSC is modeled by effective
$p\,$-wave spinless superconductor Hamiltonians [\onlinecite{Kitaev2001}-\onlinecite{Nature2016}]
\begin{align}
H^\alpha_{\rm TSC}
&=\!\sum_{i}\big[\!-\!\frac{\Delta_{\alpha,i}}{2}c_{\alpha,i+1}c_{\alpha,i}-\frac{w_{\alpha,i}}{2}c^\dag_{\alpha,i+1}c_{\alpha,i}+{\rm H.c.}\notag\\
&~~~~~~~~~~+\mu_{\alpha,i}c^\dag_{\alpha,i}c_{\alpha,i} \big] \notag\\
&=\!-i\sum^{N/2-1}_{i=1}\!\!\frac{w}{2}\gamma_{\alpha,2i}\gamma_{\alpha,2i+1}+i\sum^{N/2}_{i=1}\!\frac{\mu_{\alpha,i}}{2}\gamma_{\alpha,2i-1}\gamma_{\alpha,2i} ,
\label{TSC}
\end{align}
where $c^\dag_{\alpha,i}$ ($c_{\alpha,i}$) is the creation (annihilation) operator of electrons at site $i$ of TSC $\alpha$ ($\alpha=L, R, M)$,
$\Delta_{\alpha,i}$ and $w_{\alpha,i}$ are the pairing and hopping amplitudes respectively, and $\mu_{\alpha,i}$ is the
electron-chemical potential. The Majorana operators $\gamma_{\alpha,2i-1}=c_{\alpha,i}+c^{\dag}_{\alpha,i}$
and $\gamma_{\alpha,2i}=-i(c_{\alpha,i}-c^{\dag}_{\alpha,i})$. For simplicity, we set $\Delta_{\alpha,i}=w_{\alpha,i}=w$
hereafter. The coupling Hamiltonian between the left and middle TSCs and between the middle and right TSCs is given by
\begin{align}
H_{\rm T}(t)=&i\frac{w_{LM}(t)}{2}\gamma_{L,2N}\gamma_{M,1}+i\frac{w_{RM}(t)}{2}\gamma_{R,2N}\gamma_{M,1},
\label{H_T}
\end{align}
where $w_{L(R)M}$ is the coupling amplitude between two Majorana sites located at the left (right) and the middle TSCs.
The electron-chemical potentials of the left and right TSCs are set to be zero ($\mu_{L,i}=\mu_{R,i}=0$) so that both
TSCs are kept in topological phase. The electron-chemical potentials of the middle TSC are prepared so that $\mu_{M,1}=\mu(t)$ and $\mu_{M,i>1}\gg \Delta_{M,i}$,
namely, the electron-chemical potential at site $1$ can be individually tuned freely by a controlling gate [\onlinecite{Nature2011}],
while all other sites are kept in non-topological phase.
As a result, initially four MZMs  in the left and right TSCs
%$\tilde{\gamma}_1$, $\tilde{\gamma}_2$, $\tilde{\gamma}_3$ and $\tilde{\gamma}_4$ (see supplementary material for detail definition)
at sites $\gamma_{L,1}$, $\gamma_{L,2N}$, $\gamma_{R,1}$ and $\gamma_{R,2N}$, see Fig.~\ref{fig1}(a),
are formed the topological qubit space under consideration.
%Thus the site $1$  of the middle TSC is initially at equilibrium with the other sites
%so that $i\langle\gamma_{M,1}(t_0)\gamma_{M,2}(t_0)\rangle=0$. As a result, initially four MZMs are formed at sites $\gamma_{L,1}$,
%$\gamma_{L,2N}$, $\gamma_{R,1}$ and $\gamma_{R,2N}$ respectively (see Fig.~\ref{fig1}a).

\begin{widetext}

\begin{figure}
\includegraphics[scale=0.35]{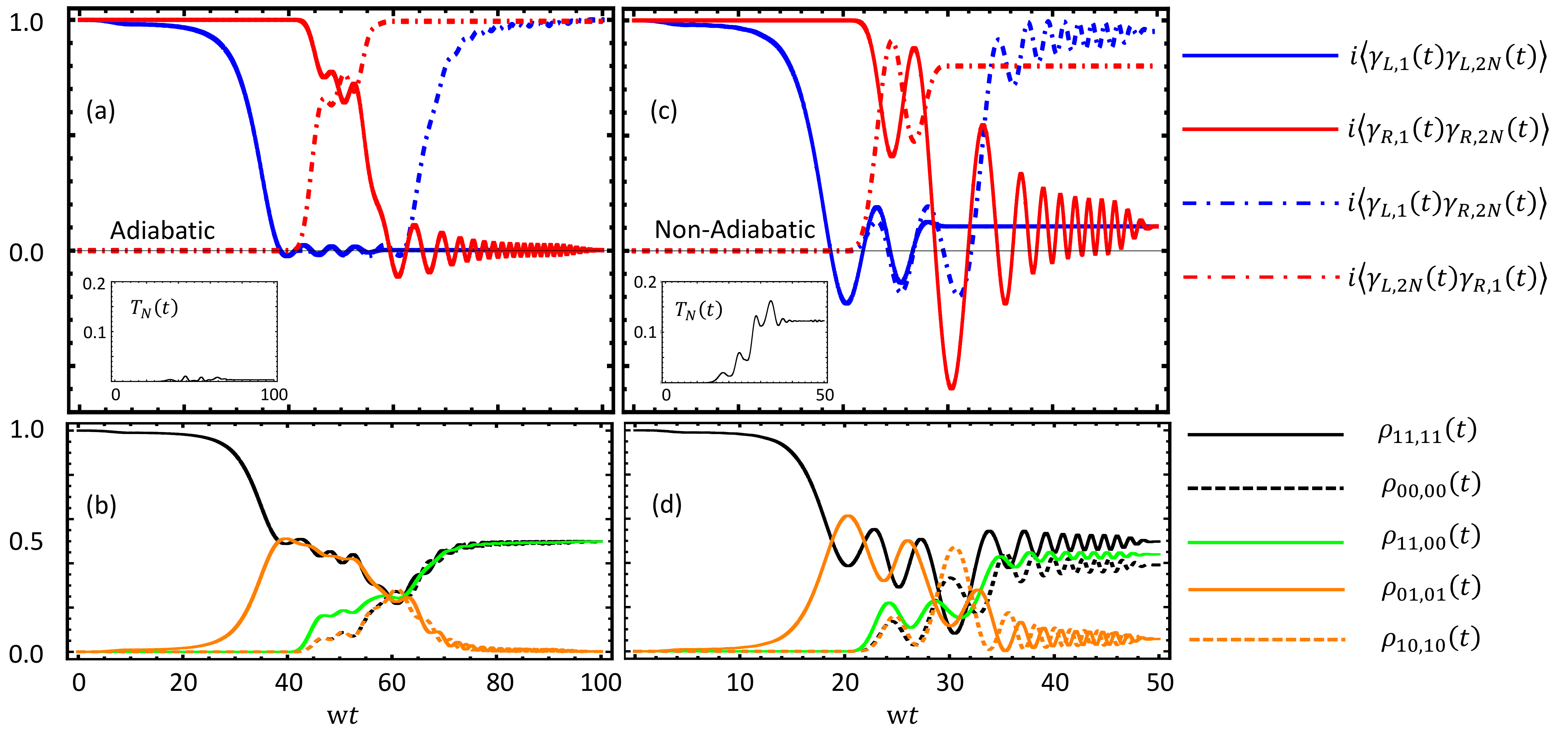}
\caption{(Colour online) (a) (Top left) Time evolution of correlation functions with total operation time $wt_T=100$. (b) (Bottom left)
Time evolution of density matrix elements with total operation time $wt_T=100$. (Inserted) Excitation number of zero-energy bogoliubons $T_N(t)$.
(c) (Top right) Time evolution of correlation functions
with total operation time $wt_T=50$. (d) (Bottom right) Time evolution of density matrix elements with total operation time
$wt_T=50$. (Inserted) Excitation number of zero-energy bogoliubons $T_N(t)$. } \label{fig2}
\end{figure}
\end{widetext}

Consider the even-parity state space of the four MZMs as the topological qubit with the basis $|00\rangle$ and $|11\rangle=b^{\dag}_{L}b^{\dag}_{R}|00\rangle$. The system is
prepared initially in one of the degenerate ground states $|11\rangle$. The empty state $|00\rangle$ is annihilated
by the left and right TSCs zero-energy bogoliubon operators $b_{L}$ and $b_{R}$, which are defined as
$b_{L}=\gamma_{L,1}+i\gamma_{L,2N}$ and $b_{R}=\gamma_{R,1}+i\gamma_{R,2N}$. In other words,
initially the MZMs at sites $\gamma_{L,1}$ ($\gamma_{R,1}$) and $\gamma_{L,2N}$ ($\gamma_{R,2N}$)
are paired to form a zero-energy bogoliubon.
The braiding is performed between the MZMs at sites $\gamma_{L,2N}$ and $\gamma_{R,2N}$, by tuning  the chemical potential
$\mu_{M,1}(t)$ of the middle TSC and the couplings $w_{LM}(t)$ and $w_{RM}(t)$ between the left TSC and the right TSC with
the middle TSC, respectively, see Fig.~\ref{fig1}b. The detailed dynamics of this braiding operation is manifested in  the correlation
functions of different Majorana sites, as shown in Fig.~\ref{fig2}. As one can see in Fig.~\ref{fig2}a, the correlation functions $i\langle\gamma_{L,1}(t)\gamma_{L,2N}(t)\rangle$
and $i\langle\gamma_{R,1}(t)\gamma_{R,2N}(t)\rangle$ gradually decrease to zero while $i\langle\gamma_{L,1}(t)\gamma_{R,2N}(t)\rangle$
and $i\langle\gamma_{L,2N}(t)\gamma_{R,1}(t)\rangle$ increase to unity.  In the end of braiding operation,
it realizes the exchange of  $\gamma_{L,2N} \leftrightarrow  \gamma_{R,2N}$, see an intuiative derivation in the Supplemental Materials \cite{SM}.
This exchange of correlations between the Majorana sites
at the left and right TSCs shows that the two zero-energy bogoliubons $b_L$ and $b_R$ become correlated to each other due to the
braiding operation.

The correlation between two zero-energy bogoliubons directly generates the Majorana qubit coherence,
as can be observed from the qubit density matrix element $\rho_{11,00}=\langle b_L(t)b_R(t)\rangle$, where the qubit is initially in the
state $|11\rangle$, and after the braiding operation, it becomes  $(|00\rangle+|11\rangle)/\sqrt{2}$, a qubit
superposition state, as shown in  in Fig.~{\ref{fig2}b}.
%Therefore, the Majorana qubit coherence is generated through the correlation between two zero-energy bogoliubons in a braiding operation.
%As it is confirmed in Fig.~{\ref{fig2}b} that the qubit state changes from $|11\rangle$ to $(|00\rangle+|11\rangle)/\sqrt{2}$ after the braiding
%process [\onlinecite{Ivanov}].
On the other hand, during the braiding operation, the closure of the even-parity qubit space is temporarily broken,
as indicated by the non-vanishing values of density matrix elements $\rho_{01,01}(t)$ and $\rho_{10,10}(t)$.
This is because one of the MZMs must be moved to the middle TSC in order to complete the
braiding operation. At the end both $\rho_{01,01}$ and $\rho_{10,10}$ vanish, indicating that the left and right TSCs
system are maintained in the even-parity state space. Also  $\rho_{00,00}=\rho_{11,11}=\rho_{11,00}$ indicating
a perfect realization of Majorana qubit coherence.

The results of Fig.~{\ref{fig2}a, b} are obtained by slowly tuning the parameter $\mu(t)$, $w_{LM}(t)$ and $w_{RM}(t)$,
as shown in Fig.~{\ref{fig1}b} with $t_T=100/w$. To examine the non-adiabatic effects of the braiding operation, the same braiding procedure is
repeated with a shorter operation time (take here, for example, $ t_T=50/w$ in Fig.~{\ref{fig1}b}). Firstly, notice the non-adiabatic characting function $T_N(t)$
(defined in the Supplementary materials \cite{SM}) which describes the number of zero-energy bogoliubon being excited
to the non-zero energy bogoliubon state due to non-adiabatic effect (see the inserted panel in Fig.~{\ref{fig2}c}).
Secondly, one can see from Fig.~\ref{fig2}c that the exchange of
MZMs between Majorana sites $\gamma_{L,2N}$ and $\gamma_{R,2N}$ cannot be completed because the controlling parameters are
changed too fast (see Fig.~\ref{fig2}c).
As a result,  the left and right bogoliubons are only partially correlated and the qubit coherence is not fully
generated  (see Fig.~{\ref{fig2}d}). The fermion parity conservation is also broken,
as shown by the non-vanished values of the old-parity matrix elements $\rho_{10,10}$ and $\rho_{01,01}$ in the
steady-state limit shown in Fig.~\ref{fig2}d, due to the non-adiabatic excitations.

%\section{Decoherence Dynamics Due to Charge Fluctuation}
\noindent {\it Decoherence Dynamics Due to Charge Fluctuations}.
In order to control the electron-chemical potential, a controlling gate is applied and we shall consider how charge fluctuations
of the controlling gate affect the braiding operation. The total Hamiltonian incorporating charge fluctuations is given by [\onlinecite{Daniel2012}]
\begin{align}
H=\sum_{\alpha=L,M,R}H^\alpha_{\rm TSC}(t)+H_{\rm T}(t)+H_G+H_C,
\end{align}
where $H_G$ is the Hamiltonian of the controlling gate, which is modeled by a non-interacting electron gas,
$H_G=\sum_{p}\epsilon_pc^\dag_pc_p$,
where $c^\dag_p$ ($c_p$) is the creation (annihilation) operator of an electron in the gate with energy $\epsilon_p$, and $H_C$ is
the coupling Hamiltonian between the TSC and the gate,
\begin{align}
H_C=\sum_{\alpha=L,M,R}\eta_{\alpha}\delta Q_{\alpha}\sum_{i}F_{\alpha,i}c^\dag_{\alpha,i}c_{\alpha,i} ,
\label{HI1}
\end{align}
in which $c^{\dag}_{\alpha,i}$ ($c_{\alpha,i}$) is the creation (annihilation) operator of an electron at site $i$ of the TSC $\alpha$
[see Eq.~(\ref{TSC})], $\eta_{\alpha}$ is the coupling parameter, $\delta Q_{\alpha}$ is the charge fluctuation of the gate and
$F_{\alpha,n}$ is a profile function which limits the range of the chain $\alpha$ affected by the gate, and one can set
 $F_{\alpha,n}$ equals to unity if site $n$ is near the TSC junction and zero otherwise. For simplicity,
we assume that all the controlling gates are identical so that we can omit the subscript $\alpha$ hereafter. The spectral density
function of the charge fluctuation of the gate is given by
\begin{align}
J_C(\omega)=B_C\exp{(-\frac{\omega^2}{8E_F\omega_c})}\frac{\omega}{1-e^{-\omega/kT}} ,
\label{JC}
\end{align}
where $E_F$ is the fermi energy of the controlling gate, $kT$ is the temperature of the gate, $\omega_c$ is a
cut-off frequency of charge fluctuations, and $B_C$ is a factor determined by the structure of the gate [\onlinecite{Daniel2012}].

We extend our previous work of the exact master equation for Majorana zero modes [\onlinecite{PRB2018}] to the topological qubit
density matrix $\rho(t)$, which is given by
\begin{align}
\frac{d}{dt}\rho(t)=&\Gamma_{L}(t,t_0)[\gamma_{L,2N}\rho(t)\gamma_{L,2N}-\rho(t)]\notag\\
&+\Gamma_{R}(t,t_0)[\gamma_{R,2N}\rho(t)\gamma_{R,2N}-\rho(t)]\notag\\
&+\Theta(t,t_0)[\gamma_{L,2N}\rho(t)\gamma_{R,2N}+\gamma_{R,2N}\rho(t)\gamma_{L,2N}]\notag\\
&+\Lambda(t,t_0)[\gamma_{L,2N}\gamma_{R,2N}\rho(t)+\rho(t)\gamma_{R,2N}\gamma_{L,2N}]\notag\\
&+\tilde{\Lambda}(t,t_0)[\gamma_{L,2N}\gamma_{R,2N}\rho(t)-\rho(t)\gamma_{R,2N}\gamma_{L,2N}].
\label{MasterEq}
\end{align}
The time-dependent coefficients $\Gamma_{L(R)}(t,t_0)$, $\Theta(t,t_0)$, $\Lambda(t,t_0)$ and $\tilde{\Lambda}(t,t_0)$ are given in the Supplementary materials.

\begin{widetext}

\begin{figure}
\includegraphics[scale=0.35]{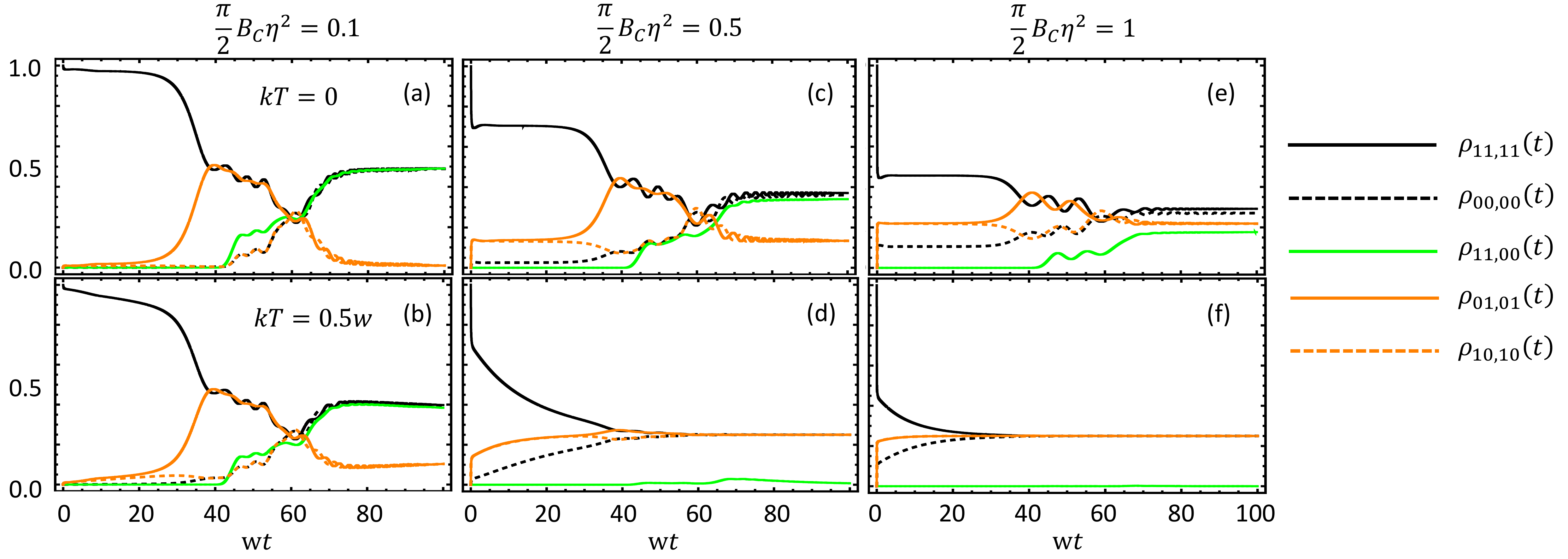}
\caption{(Colour online) Time evolution of density matrix elements with (a) (top left) $\frac{\pi}{2}B_C\eta^2=0.1$,
$kT=0$, (b) (bottom left) $\frac{\pi}{2}B_C\eta^2=0.1$, $kT=0.5w$, (c) (top middle) $\frac{\pi}{2}B_C\eta^2=0.5$, $kT=0$,
(d) (bottom middle) $\frac{\pi}{2}B_C\eta^2=0.5$, $kT=0.5w$, (e) (top right) $\frac{\pi}{2}B_C\eta^2=1$, $kT=0$,
(f) (bottom right) $\frac{\pi}{2}B_C\eta^2=1$, $kT=0.5w$, with initial qubit state $|11\rangle$.}  \label{fig3}
\end{figure}
\end{widetext}

We perform the same braiding procedure as in the previous section under the charge fluctuations of the gates.
Fig.~\ref{fig3} shows the decoherence dynamics of the TSC-system density matrix elements at various coupling regimes
for both the zero temperature and finite temperature ($kT=0.5w$). First of all, MZMs at sites $\gamma_{L,2N}$
and $\gamma_{R,2N}$ are subjected to the charge fluctuation, leading to the dissipation of the corresponding
bogoliubon correlations. As we have mentioned above, the qubit coherence is generated through the formation
of bogoliubon correlations between the two TSCs so that the dissipation of MZMs will lead to the destruction of
qubit coherence, as illustrated by $\rho_{11,00}(t)$ in Fig.~\ref{fig3}. At zero temperature,
the MZMs are partially protected by the localized bound states due to the particle-hole symmetry [\onlinecite{PRB2018}],
part of the bogoliubon correlations can still be formed in the braiding operation so that the qubit
coherence is partially generated and preserved (see Fig.~\ref{fig3}a, c, e). In contrast,
at finite temperature, the localized bound states do not exist and the fluctuation-affected MZMs
eventually decay to zero [\onlinecite{PRB2018}]. In particular, in the weak coupling regime ($\frac{\pi}{2}B_C\eta^2<0.5$),
the MZMs decay comparatively slower so that the bogoliubon correlations can still be formed in the braiding process,
thus the qubit coherence can be partially generated before it decays to zero (see Fig.~\ref{fig3}b, d).
However, in the strong coupling regime ($\frac{\pi}{2}B_C\eta^2=1$), the MZMs decay so quickly that the
bogoliubon correlations cannot be formed during the braiding process. Therefore, qubit coherence cannot
be generated at all in the strong coupling regime (see Fig.~\ref{fig3}f).

Secondly, the fermion parity conservation and thus the closure of the qubit state space are also broken by
charge fluctuations. The qubit density matrix elements $\rho_{01,01}(t)$ and $\rho_{10,10}(t)$ describe
the bogoliubon occupation number in the odd-parity state space, their non-vanishing values indicate the breakdown of the
parity-conserved qubit space. At zero temperature, the localized bound states suppress the breakdown
of the fermion parity conservation (see Fig.~\ref{fig3}a, c, e). While at finite temperature, two bogoliubons
(four MZMs) eventually evolve into a totally mixed state (see Fig.~\ref{fig3}b, d, f).

In conclusion, we have studied the decoherence dynamics of a Majorana qubit under braiding operation in a T-juction-TSC system with the exact master equation.
During the braiding process, two MZMs in the right and left sides of two TSCs exchange so that two zero-energy bogoliubons become correlated to each other.
The Majorana qubit coherence is generated through this correlation formation between two bogoliubons in different TSCs.
We show that non-adiabatic decoherence will damage the qubit coherence because of the incomplete formation of bogoliubon correlation.
It also breaks the fermion parity conservation because of the non-adiabatic excitations.
When charge fluctuations induced from the controlling gate are taken into account, the fluctuation-affected MZMs in both
TSCs will dissipate.
%At zero temperature, due to the particle-hole symmetry, there are localized bound states that partially
%protect the MZMs so the qubit coherence can be partially generated and preserved in the braiding operation.
%At finite temperature, localized bound state no longer exists so that the MZMs cannot be protected anyway.
As a result, in the weak coupling regime, the qubit coherence is only partially generated in the beginning of the braiding operation,
and then the coherence decays away eventually. While in the strong coupling regime, the qubit coherence cannot be generated at all.
Also we show that
%at zero temperature the localized bound states can only suppress the breakdown of fermion parity conservation, while
at finite temperature, the four MZMs eventually become a totally mixed state so that
the information encoded in the Majorana qubit is completely lost in topological quantum computations.
These results show that the advantages of Majorana qubits from the non-locality and the fermion parity conservation
are destroyed by local perturbations.

%\section*{Acknowledgement}
We acknowledges the support from the Ministry of Science and Technology of Taiwan under
Contract Nos.~NSC-105-2112-M-006-008-MY3 and NSC-108-2112-M-006-009-MY3.

\end{document}

\title{Supplementary materials}
\author{Hon-Lam Lai}
\affiliation{Department of Physics and Center for Quantum
information Science, National Cheng Kung University, Tainan 70101,
Taiwan}
\author{Wei-Min Zhang}
\email{wzhang@mail.ncku.edu.tw}
\affiliation{Department of Physics
and Center for Quantum information Science, National Cheng Kung
University, Tainan 70101, Taiwan}

\pacs{72.10.Bg, 73.63.-b, 85.30.Mn} \maketitle

\section{Majorana braiding and the master equation}
The master equation of the TSC-system is a generalization of the result we have obtained in our previous work [\onlinecite{PRB2018}],
\begin{align}
\frac{d}{dt}\rho(t)=&\Gamma_{L}(t,t_0)[\gamma_{L,2N}\rho(t)\gamma_{L,2N}-\rho(t)]\notag\\
&+\Gamma_{R}(t,t_0)[\gamma_{R,2N}\rho(t)\gamma_{R,2N}-\rho(t)]\notag\\
&+\Theta(t,t_0)[\gamma_{L,2N}\rho(t)\gamma_{R,2N}+\gamma_{R,2N}\rho(t)\gamma_{L,2N}]\notag\\
&+\Lambda(t,t_0)[\gamma_{L,2N}\gamma_{R,2N}\rho(t)+\rho(t)\gamma_{R,2N}\gamma_{L,2N}]\notag\\
&+\tilde{\Lambda}(t,t_0)[\gamma_{L,2N}\gamma_{R,2N}\rho(t)-\rho(t)\gamma_{R,2N}\gamma_{L,2N}],
\end{align}
where
\begin{align}
\Gamma_{L}(t,t_0)=-\big[&\dot{\bm{U}}(t,t_0)\bm{U}^{-1}(t,t_0)\big]_{11}\notag\\
\Gamma_{R}(t,t_0)=-\big[&\dot{\bm{U}}(t,t_0)\bm{U}^{-1}(t,t_0)\big]_{22}\notag\\
\Theta(t,t_0)=-\frac{1}{2}\{&\big[\dot{\bm{U}}(t,t_0)\bm{U}^{-1}(t,t_0)\big]_{12}\notag\\
&+\big[\dot{\bm{U}}(t,t_0)\bm{U}^{-1}(t,t_0)\big]_{21}\}\notag\\
\Lambda(t,t_0)=-\frac{1}{2}\{&\big[\dot{\bm{U}}(t,t_0)\bm{U}^{-1}(t,t_0)\big]_{21}\notag\\
&-\big[\dot{\bm{U}}(t,t_0)\bm{U}^{-1}(t,t_0)\big]_{12}\},
\end{align}
where the function $\tilde{\Lambda}(t)$, of which the definition will be given later, is dependent on the initial condition
$i\langle\gamma_{M,1}(t_0)\gamma_{M,2}(t_0)\rangle$ and $\bm{U}(t,t_0)$ is the Green function matrix which is defined by
\begin{align}
\bm{U}(t,t_0)=
\begin{pmatrix}
i\langle\gamma_{L,1}(t_0)\gamma_{L,2N}(t)\rangle & i\langle\gamma_{R,1}(t_0)\gamma_{L,2N}(t)\rangle \\
i\langle\gamma_{L,1}(t_0)\gamma_{R,2N}(t)\rangle & i\langle\gamma_{R,1}(t_0)\gamma_{R,2N}(t)\rangle
\end{pmatrix}
\end{align}
with the initial condition $i\langle\gamma_{L,1}(t_0)\gamma_{L,2N}(t_0)\rangle=i\langle\gamma_{R,1}(t_0)\gamma_{R,2N}(t_0)\rangle=1$.

To show intuitively the braiding manipulation,
we first rewrite the Hamiltonian of the TSC system without charge fluctuation, which is the combination of  Eq.~(1) and (2) in the main text
\begin{align}
\sum_{\alpha}H^\alpha_{\rm TSC}(t)+H_{\rm T}(t)=&\!\sum_kE_{k}b^{\dag}_{M,k}b_{M,k}+\!\sum_{\alpha=L,R;k}w b^{\dag}_{\alpha, k}b_{\alpha, k}\notag\\
&+E(t)\tilde{b}^{\dag}_{M}(t)\tilde{b}_{M}(t).
\end{align}
The first term of the Hamiltonian corresponds to the bogoliubon modes of the middle TSC in non-topological phase
and do not contribute to the braiding dynamics.
The second term corresponds to the bogoliubon modes of the left and right TSCs in topological phase. The last term describes the non-adiabatic excitations
of the system during the braiding operation. The operator $\tilde{b}^{\dag}_{M}(t)$ creates bogoliubons with energy $E(t)=\sqrt{\mu^2(t)+
w^2_{LM}(t)+w^2_{RM}(t)}$ is a mixture of the bogoliubon $b_{M,1}=\gamma_{M,1}+i\gamma_{M,2}$ of the middle TSC with other two zero-energy
bogoliubons $b_{L}=\gamma_{L,1}+i\gamma_{L,2N}$ and $b_{R}=\gamma_{R,1}+i\gamma_{R,2N}$ in the left and right TSCs
through the couplings between them given by Eq.~(2) in the main text.
Explicitly,
%are absent in the above Hamiltonian. The time-dependent operators are given by
\begin{widetext}
\begin{subequations}
\begin{align}
& \left\{
\begin{array}{ll}
\tilde{b}_{L}(t)=\frac{1}{2}\gamma_{L,1}+\frac{i}{2\sqrt{\mu^2(t)+w_{LM}^2(t)}}[\mu(t)\gamma_{L,2N}+w_{LM}(t)\gamma_{M,2}],\\& \\
\tilde{b}_R(t)=\frac{1}{2}\gamma_{R,1}+\frac{i}{2}\gamma_{R,2N},~~~~~~~~~~~~~~~~~~~~~~~~~~~~~~& t_0 \leq t<t_1\\ &\\
\tilde{b}_M(t)=\frac{1}{2}\gamma_{M,1}+\frac{i}{2\sqrt{\mu^2(t)+w_{LM}^2(t)}}[w_{LM}(t)\gamma_{L,2N}-\mu(t)\gamma_{M,2}],  &
\end{array}
\right.  \\  \notag \\
& \left\{
\begin{array}{ll}
\tilde{b}_L(t)=\frac{1}{2}\gamma_{L,1}+\frac{i}{2}\gamma_{M,2},~~~~~~~~~~~~~~~~~~~~~~~~~~~~~~~~  & \\ ~ &~ \\
\tilde{b}_R(t)=\frac{1}{2}\gamma_{R,1}+\frac{i}{2\sqrt{w_{LM}^2(t)+w_{RM}^2(t)}}[-w_{RM}(t)\gamma_{L,2N}+w_{LM}(t)\gamma_{R,2N}], &  t_1\leq t\leq t_2 \\~ &~  \\
\tilde{b}_M(t)=\frac{1}{2}\gamma_{M,1}+\frac{i}{2\sqrt{w_{LM}^2(t)+w_{RM}^2(t)}}[w_{LM}(t)\gamma_{L,2N}+w_{RM}(t)\gamma_{R,2N}] ,&
\end{array}
\right. \\  \notag \\
& \left\{
\begin{array}{ll}
\tilde{b}_L(t)=\frac{1}{2}\gamma_{L,1}+\frac{i}{2\sqrt{\mu^2(t)+w_{RM}^2(t)}}[\mu(t)\gamma_{R,2N}+w_{RM}(t)\gamma_{M,2}], & \\  & \\
\tilde{b}_R(t)=\frac{1}{2}\gamma_{R,1}-\frac{i}{2}\gamma_{L,2N},~~~~~~~~~~~~~~~~~~~~~~~~~~~~~~ & t>t_2 \\ &  \\
\tilde{b}_M(t)=\frac{1}{2}\gamma_{M,1}+\frac{i}{2\sqrt{\mu^2(t)+w_{RM}^2(t)}}[w_{RM}(t)\gamma_{R,2N}-\mu(t)\gamma_{M,2}]. &
\end{array}. \label{tdmajora}
\right.
\end{align}
\label{b_tilde}
\end{subequations}
\end{widetext}
Thus, the time-dependent MZM operators during the braiding operations are given by
\begin{subequations}
\label{gamma_tilde}
\begin{align}
&\tilde{\gamma}_{L,1}(t)= \tilde{b}_L (t)+\tilde{b}^\dag_L(t) =b_L +b^\dag_L  = \gamma_{L,1} , \\
&\tilde{\gamma}_{L,2N}(t)=-i\big[\tilde{b}_L(t)-\tilde{b}^\dag_L (t)\big] , \\
&\tilde{\gamma}_{R,1}(t)=\tilde{b}_R(t)+\tilde{b}^\dag_R(t)=b_R +b^\dag_R  = \gamma_{R,1} , \\
&\tilde{\gamma}_{R,2N}(t)=-i\big[\tilde{b}_R(t)-\tilde{b}^\dag_R(t)\big] ,
\end{align}
\end{subequations}
Note that operators $\tilde{\gamma}_{L,1}(t)$ and $\tilde{\gamma}_{R,1}(t)$ are time-independent because
they are located at the end of the two sides of the L-R TSCs so that the braiding operations on the sites $\{L,2N\}$ and $\{R,2N\}$ of the T-junction
will not locally affect these two Majorana modes, see Fig. 1(a). It is clear that in the end of braiding, we obtained from Eq.~(\ref{tdmajora}) that
\begin{align}
\tilde{\gamma}_{L,2N}(t) \longrightarrow \gamma_{R,2N} ~~~,~~~~  \tilde{\gamma}_{R,2N}(t) \longrightarrow -\gamma_{L,2N}  ,
\end{align}
which shows a realization of braiding operations. We can also define the non-adiabatic excitation number $T_N(t)$ of the zero-energy bogoliubon as
\begin{align}
T_N(t)&=\big[1-\langle\tilde{b}^{\dag}_{L}(t)\tilde{b}_{L}(t)\rangle\big]+\big[1-\langle\tilde{b}^{\dag}_{R}(t)\tilde{b}_{R}(t)\rangle\big]\notag\\
&=\langle\tilde{b}^{\dag}_M(t)\tilde{b}_M(t)\rangle-\langle\tilde{b}^{\dag}_M(t_0)\tilde{b}_M(t_0)\rangle,
\end{align}
to characterize the non-adiabatic braiding operation.

Next, consider the charge fluctuation of the controlling gate on the real-time evolution of the MZMs in braiding operations.
Using Eq.~(\ref{b_tilde}) in the main text, we rewrite the Hamiltonian (4) by keeping only terms that are linearly proportional
to $\gamma_{L,2N}$ and $\gamma_{R,2N}$ [\onlinecite{Daniel2012}],
\begin{align}
H_C=&-i\eta\delta Q\sum_{\alpha;k\neq 0}X_{\alpha k}\big[\gamma_{\alpha,2N}(b_{\alpha k}+b^{\dag}_{\alpha k})\big]\notag\\
&-i\eta\delta Q\gamma_{M,1}\gamma_{M,2},
\end{align}
where $\alpha=L,R$.

The time evolution of the two Majorana operators $\gamma_{L,2N}(t)$ and $\gamma_{R,2N}(t)$ can be determined by the
Heisenberg equation of motion after eliminating the Majorana operators $\gamma_{M,1}$, $\gamma_{M,2}$ and all the finite-energy
bogoliubons $b_{Lk}$, $b_{Rk}$,
\begin{align}
\frac{d}{dt}
\begin{pmatrix}
\gamma_{L,2N}(t)\\
\gamma_{R,2N}(t)
\end{pmatrix}
=-&\!\int^t_{t_0}\!\!d\tau\!
\begin{pmatrix}
\mathcal{G}_{LL}(t,\tau) & \mathcal{G}_{LR}(t,\tau)\\
\mathcal{G}_{RL}(t,\tau) & \mathcal{G}_{RR}(t,\tau)
\end{pmatrix}\!
\begin{pmatrix}
\gamma_{L,2N}(\tau)\\
\gamma_{R,2N}(\tau)
\end{pmatrix}\notag\\
&+\bm{f}(t),
\label{EOM}
\end{align}
where $\bm{f}(t)$ is a noise operator related to the initial operators $\gamma_{M,1}(t_0)$, $\gamma_{M,2}(t_0)$, $b_{Lk}(t_0)$
and $b_{Rk}(t_0)$. The integral kernel matrix elements are obtained by eliminating the Majorana operator $\gamma_{M,1}$ and $\gamma_{M,2}$,
\begin{align}
\mathcal{G}&_{\alpha\beta}(t,\tau)=w_{\alpha}(t)w_{\beta}(\tau)F_{\rm{CF}}(t,\tau)\notag\\
&+\delta_{\alpha\beta}\sum_{k}X^2_{\alpha k}
{\rm{Re}}\big[\eta^2\langle\delta Q(t)\delta Q(\tau)\rangle e^{-iw(t-\tau)}\big]
\end{align}
where the fator $F_{\rm{CF}}(t,\tau)$ is induced by the charge fluctuation of braiding operations, which satisfies
\begin{align}
\frac{d}{d\tau}F_{\rm{CF}}(t,\tau)=&\int^t_{\tau}d\tau'\mu(t)\mu(\tau')F_{\rm{CF}}(t,\tau')\notag\\
&+\eta^2\int^t_{\tau}d\tau'{\rm{Re}}\langle\delta Q(t)\delta Q(\tau')\rangle F_{\rm{CF}}(t,\tau')
\end{align}
and $F_{\rm{CF}}(t,t)=1$.

With Eq.~(\ref{b_tilde}) and Eq.~(\ref{gamma_tilde}), one can write
\begin{align}
\begin{pmatrix}
\gamma_{L,2N}(t)\\
\gamma_{R,2N}(t)
\end{pmatrix}
&=\bm{U}(t,t_0)
\begin{pmatrix}
\gamma_{L,2N}(t_0)\\
\gamma_{R,2N}(t_0)
\end{pmatrix}\notag\\
&+\bm{M}(t,t_0)
\begin{pmatrix}
\gamma_{M,1}(t_0)\\
\gamma_{M,2}(t_0)
\end{pmatrix}
+\bm{F}(t),
\end{align}
where
\begin{widetext}
\begin{align}
\bm{M}(t,t_0)=
\!\int^t_{t_0}\!\!d\tau\bm{U}(t,\tau)
\begin{pmatrix}
w_L(\tau)F_{\rm{CF}}(\tau,t_0) & w_L(\tau)\int^{\tau}_{t_0}d\tau'F_{\rm{CF}}(\tau,\tau')\mu(\tau')\\
w_R(\tau)\int^{\tau}_{t_0}d\tau'F_{\rm{CF}}(\tau,\tau')\mu(\tau') & w_R(\tau)F_{\rm{CF}}(\tau,t_0)
\end{pmatrix}
\end{align}
\end{widetext}
and $\bm{F}(t)=\!\int^t_{t_0}\!\!d\tau\bm{U}(t,\tau)\bm{f}(\tau)$. The Green function $\bm{U}(t,t_0)$ can be obtained by
solving the differential-integral equation (\ref{EOM}) and the time-dependent coefficient $\tilde{\Lambda}(t,t_0)$ of the master equation is given by
\begin{align}
\tilde{\Lambda}(t,t_0)=-\frac{1}{2}\langle\gamma_{M,1}(t_0)\gamma_{M,2}(t_0)\rangle
&\frac{d}{dt}\big[\bm{M}_{11}(t,t_0)\bm{M}_{22}(t,t_0)\notag\\
&-\bm{M}_{12}(t,t_0)\bm{M}_{21}(t,t_0)\big].
\end{align}